\documentclass[aps,prc,twocolumn,superscriptaddress,showpacs,showkeys]{revtex4-1}
\usepackage{graphicx}
\begin{document}
\newcommand{\Jpsi}{J/\psi}
\newcommand{\pT}{p_{T}}

\title{Charged current deep inelastic scattering of $\nu_{\mu}$ off $^{56}$Fe}

\author{Deepika Grover}
\email{dgroverbhu@gmail.com}
\affiliation{Department of Physics, Institute of Science, Banaras Hindu University, Varanasi 221005, India}
\author{Kapil Saraswat}
\email{kapilsaraswatbhu@gmail.com}
\affiliation{Department of Physics, Institute of Science, Banaras Hindu University, Varanasi 221005, India}
\author{Prashant Shukla}
\email{pshuklabarc@gmail.com}
\affiliation{Nuclear Physics Division, Bhabha Atomic Research Centre, Mumbai 400085, India}
\affiliation{Homi Bhabha National Institute, Anushakti Nagar, Mumbai 400094, India}
\author{Venktesh Singh}
\email{venkaz@yahoo.com}
\affiliation{Department of Physics, Institute of Science, Banaras Hindu University, Varanasi 221005, India}
\date{\today}

\begin{abstract}
  In this paper, we study charged current deep inelastic scattering of muon neutrinos off $^{56}$Fe nuclei using Hirai, Kumano and Saito model. The LHA Parton Distribution Functions (PDFs) - CT10 are used to describe the partonic content of hadrons. Modification of PDFs inside the nuclei is done using EPPS16 parameterization at next-to-leading order. Target mass correction has also been incorporated in the calculations. We calculate the structure functions ($F_{2}(x,Q^{2})$ and $xF_{3}(x,Q^{2})$), the ratios ($R_{2}(x,Q^{2}) = \frac{F^{^{56}Fe}_{2}}{F^{Nucleon}_{2}}$ and $R_{3}(x,Q^{2}) = \frac{F^{^{56}Fe}_{3}}{F^{Nucleon}_{3}}$) and the differential cross sections of $\nu_{\mu}$ deep inelastic scattering off a nucleon and $^{56}$Fe nuclei. We compare the obtained results with measured experimental data. The present theoretical approach gives a good description of data.
\end{abstract}

\maketitle

\section{Introduction}
 In the standard model of physics, neutrinos are elementary particles with no electric charge, no magnetic moment, half integral spin and zero mass. However, several neutrino oscillation experiments~\cite{Ahn:2002up,Aliu:2004sq,Ahn:2006zza,Ashie:2005ik,Takeuchi:2011aa,Adamson:2016tbq,Adamson:2016xxw,Adamson:2017qqn,Adamson:2017gxd,Kumar:2017sdq} across the globe have confirmed that neutrinos oscillate from one flavor to another, leading to a small but non-zero neutrino mass and the possibility to go beyond standard model. Being electrically neutral, neutrinos rarely interact with matter via weak force. Neutrino interactions are classified into two categories: Charged Current (CC) interactions via the exchange of $W^{+}/W^{-}$ boson and Neutral Current (NC) interactions via the exchange of $Z$ boson. There are many neutrino scattering processes such as quasi elastic scattering (QES)~\cite{Grover:2018out}, resonance pion production (RES)~\cite{Saraswat:2016kln} and deep inelastic scattering (DIS), at various neutrino energies, for a review see Ref.~\cite{Formaggio:2013kya}. Low neutrino energies are dominated by QES whereas RES dominates at medium neutrino energies. As the neutrino energies become larger, DIS becomes more and more dominant~\cite{Lipari:2002at}. In this scattering process, highly energetic neutrino scatters off a quark in the nucleon producing a corresponding lepton and many hadrons are produced.
\begin{equation}
\nu_{\mu} + N \rightarrow \mu^{-} + X ,~~~~~(CC)
\end{equation}
\begin{equation}
\nu_{\mu} + N \rightarrow \nu_{\mu} + X .~~~~~(NC)
\end{equation}

DIS is an important experimental tool for studying the hadronic matter where the final state particles produced in the scattering are analyzed to probe hadronic properties. Several experiments planned worldwide such as NuTeV~\cite{Tzanov:2005kr}, CHORUS~\cite{Onengut:2005kv}, NOMAD~\cite{Wu:2007ab}, MINOS~\cite{Bhattacharya}, MINERvA~\cite{Mousseau:2016snl} etc. have analyzed neutrino deep inelastic scattering off different targets to measure differential and integrated cross sections and structure functions. A review of the results from various experiments probing neutrino deep inelastic scattering in presented in Ref.~\cite{Tzanov:2009zz}. 
 
 To describe the partonic content of the hadron, precise parton distribution functions (PDFs) are required. These PDFs are produced by several different groups such as MRST~\cite{Martin:2009iq,Martin:2009bu,Martin:2010db}, CTEQ~\cite{Nadolsky:2008zw},
Alekhin~\cite{Alekhin:2013nda,Alekhin:2015cza}, ZEUS~\cite{Chekanov:2002pv} etc. PDFs are derived from fitting DIS and related hard scattering data using parameterization at low $Q^{2}_{0}(1-7 ($GeV/c$)^{2})$ and evolving these to higher $Q^{2}$. These PDFs are presented as grids in $x-Q^{2}$ with codes given by PDF authors. LHAPDF (The Les Houches Accord PDFs) provides C++ code to these PDFs with interpolation grid build into the PDFLIB~\cite{whalley}. We use LHAPDF (CT10)~\cite{Lai:2010vv} parton distribution functions.

 In this work, we study charged current $\nu_{\mu}$ - nucleon and $\nu_{\mu}$ - $^{56}$Fe deep inelastic scattering using Hirai, Kumano and Saito model~\cite{hirai}. Calculations of the structure functions ($F_{2}(x,Q^{2})$ and $xF_{3}(x,Q^{2})$), the ratios ($R_{2}(x,Q^{2}) = \frac{F^{^{56}Fe}_{2}}{F^{Nucleon}_{2}}$ and $R_{3}(x,Q^{2}) = \frac{F^{^{56}Fe}_{3}}{F^{Nucleon}_{3}}$) and the differential cross sections are presented and compared with the measured experimental data.


\section{Formalism for deep inelastic $\nu - N$ and $\nu - A$ scattering}

\begin{figure}[htp]
\centering
\begin{tabular}{cc}
\includegraphics[width=80mm]{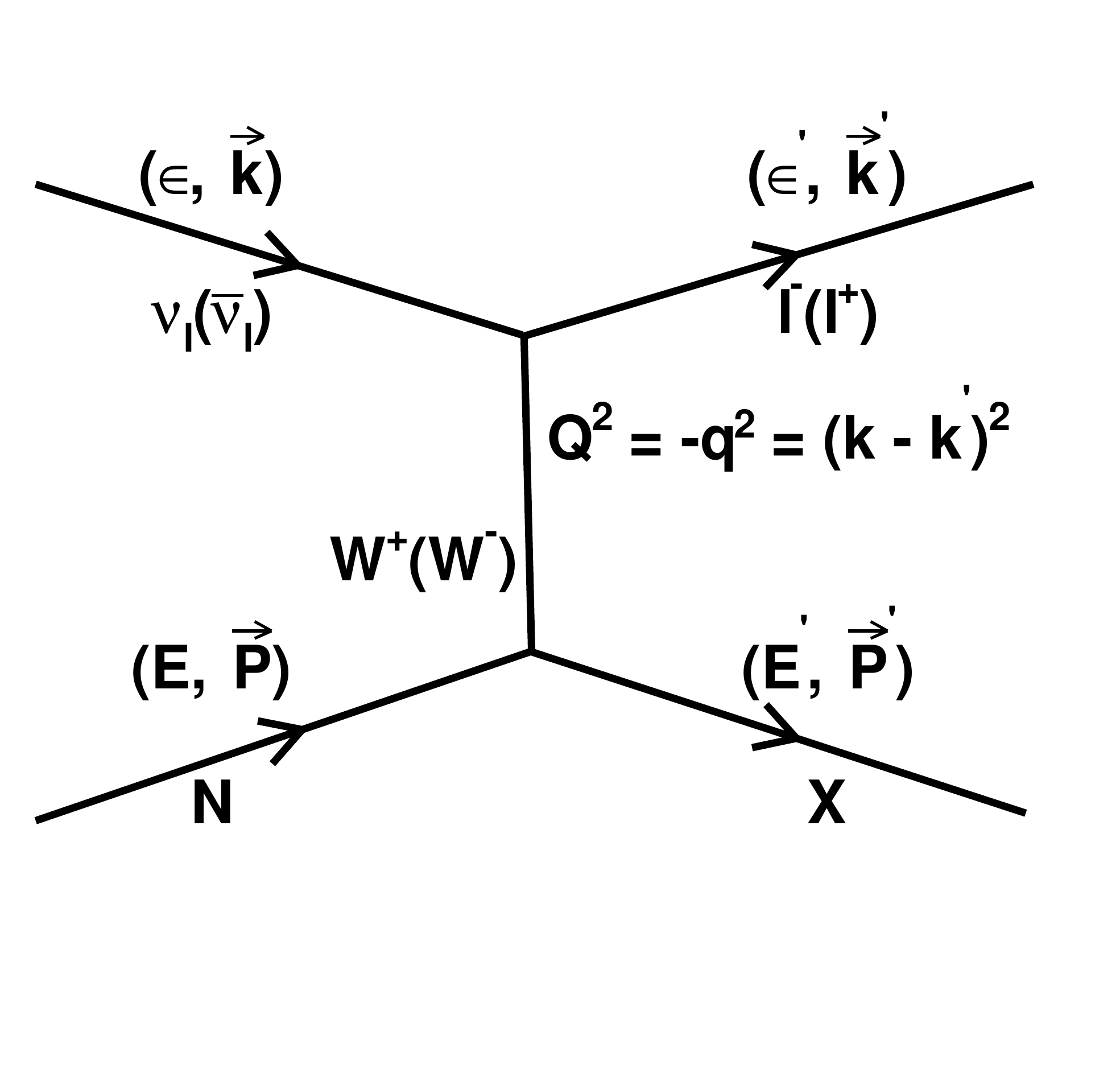}\\
\end{tabular}
\caption{Charged current $\nu - N$ deep inelastic scattering.}
\label{DIS_Feynman_Diagram}
\end{figure}

 The neutrino (anti-neutrino) - nucleon deep inelastic scattering process is:
\begin{eqnarray}
\nu_{l}(k) + N(P) \rightarrow l^{-}(k') + X(P') ,      \\
\bar\nu_{l}(k) + N(P) \rightarrow l^{+}(k') + X(P') .
\end{eqnarray}
where a neutrino or anti-neutrino with 4-momentum $k=(\epsilon,\vec{k})$ scatters off a nucleon $N$ with 4-momentum $P^{\mu}=(E,\vec{P})$ and $E=\sqrt{M^{2}+\vec{P}^{2}}$. 
The outgoing lepton $l^{-}$ or $l^{+}$ (not neutrino) has 4-momentum $k'=(\epsilon^{'},\vec{k'})$. The 
hadronic final state $X$ is left with a 4-momentum $P'=(E',\vec{P'})$. The schematic diagram of charged current $\nu - N$ deep inelastic scattering is shown in figure \ref{DIS_Feynman_Diagram}.

  Neutrino CC (Charged Current) interactions with nucleon are described by the 
matrix elements~\cite{hirai}:
\begin{eqnarray}
M&=&\frac{G_F}{\sqrt 2} \, \frac{M_{W}^{2}}{M_{W}^{2} + Q^2} \, \,
 \bar u(k',\lambda ') \, \nonumber \\
&& \gamma ^\mu (1 - \gamma_5) \, u(k,\lambda )  
\left\langle  X \left| J_\mu ^{CC} (0) \right|  {P,\lambda _N } \right\rangle .
\end{eqnarray}
where $G_F$ is the Fermi coupling constant,
$M_W$ is the $W$ mass, $Q^2$ is given by $Q^{2}=-q^{2}=(k-k')^{2}$  with the four-momentum transfer $q$,
$k$ ($\lambda$) and $k'$ ($\lambda'$) indicate initial and final lepton momenta (spins), 
$P$ ($\lambda_N$) is the nucleon momentum (spin) and $J_\mu ^{CC} (0)$ is the weak charged 
current (CC) of the nucleon. The absolute value square $|M|^2$ is calculated with an average
over the nucleon spin for obtaining the differential cross section.

 The neutrino (anti-neutrino) - nucleon charged current differential scattering cross section is 
defined as~\cite{hirai}:
\begin{eqnarray}
\frac{d^2\sigma^{\nu(\bar\nu)}_{CC}}{dxdy}&=&\frac{G_F^2 M_N E_{\nu}}{\pi} \, \Bigg(\frac{M^2_{W}}{M^2_{W} + Q^2}\Bigg)^{2} \,
\Bigg[ F_{1}(x,Q^{2}) xy^2  \nonumber \\
&&+ ~F_{2}(x,Q^{2})
\left( {1 - y - \frac{{M_N xy}}{{2E_{\nu}}}} \right) \nonumber \\
&&\pm ~F_{3}(x,Q^{2}) xy\left( {1 - \frac{y}{2}} \right) \Bigg] .
\end{eqnarray}
where $\pm$ indicates $+$ for $\nu$ and $-$ for $\bar \nu$,
$x$ is the Bjorken scaling variable defined as $x=Q^2/(2\, P\cdot q)$,
$y$ is the inelasticity defined as $y=P\cdot q/(P\cdot k)$, hence $Q^{2} = 2M_{N}E_{\nu}xy$, 
$E_{\nu}$ is the neutrino-beam energy and $M_N$ is the nucleon mass.
$F_{1}(x,Q^{2})$, $F_{2}(x,Q^{2})$ and $F_{3}(x,Q^{2})$ are the dimensionless structure functions. The Bjorken variable $x$ and the inelasticity $y$ are in the range $0 \leq x,y \leq 1$.

In the Bjorken limit of scaling in the asymptotic region i.e. $Q^{2} \rightarrow \infty$, $x$ is 
finite. The structure functions $F_{i}(x,Q^{2})$ are not $Q^{2}$ dependent and depend only 
on $x$, and satisfy the Callan-Gross relation~\cite{callan}:
\begin{equation}
F_{2}(x)=2xF_{1}(x) .
\end{equation}
  Using Callan-Gross relation, the differential cross section can be expressed in terms of $F_{2}$ 
and $F_{3}$. In Quark Parton Model (QPM), $F_{2}(x,Q^{2})$ and $F_{3}(x,Q^{2})$ are determined in terms of PDFs for quarks and 
anti-quarks.

The structure function $F_{3}(x,Q^{2})$ is defined as:
\begin{eqnarray}\label{}
F^{N}_{3}(x,Q^{2})&=& \Bigg(u(x,Q)-\bar u(x,Q)+d(x,Q) \nonumber \\
&&-~\bar d(x,Q)+s(x,Q)-\bar s(x,Q) \nonumber \\
&&+~c(x,Q)-\bar c(x,Q)+b(x,Q) \nonumber \\
&&-~\bar b(x,Q)+t(x,Q)-\bar t(x,Q)\Bigg) .    
\end{eqnarray}

The structure function $F_{2}(x,Q^{2})$ is defined as:
\begin{eqnarray}\label{}
F^{N}_{2}(x,Q^{2})&=&~x\Bigg(u(x,Q)+\bar u(x,Q)+d(x,Q) \nonumber \\
&&+~\bar d(x,Q)+s(x,Q)+\bar s(x,Q) \nonumber \\
&&+~c(x,Q)+\bar c(x,Q)+b(x,Q) \nonumber \\
&&+~\bar b(x,Q)+t(x,Q)+\bar t(x,Q)\Bigg) .       
\end{eqnarray}

\section{Nuclear modifications}
\label{section_nuc_modifications}
  There are some nuclear effects~\cite{Armesto:2006ph} in neutrino - nucleus deep inelastic scattering process. These effects were first pointed out in 1982 by the EMC collaboration at CERN, where they measured the ratio of iron ($F^{A}_{2}(x,Q^{2})$) to deuterium ($F^{D}_{2}(x,Q^{2})$) structure functions, and found the results to be different from unity~\cite{Aubert:1983xm}. Several efforts since then have been made to explore these effects in neutrino - nucleus DIS process with the conclusion that for Bjorken variable $x < 0.1$, the ratio is suppressed and the suppression increases with increase in the mass number of the target nucleus. This suppression is called shadowing effect. For $0.1 < x < 0.3$, the ratio is more than unity. This increase in the ratio is called anti-shadowing effect. For $0.3 < x < 0.8$, the ratio is again suppressed and this suppression is called EMC effect. For $x > 0.8$, the ratio increases rapidly and this rapid increase in the ratio is due to Fermi motion effect. For a review, see Ref.~\cite{Arneodo:1992wf}.
  
  The neutrino (anti-neutrino)-nucleus charged current differential scattering cross section
is defined as~\cite{haider}:
\begin{eqnarray}\label{Diff_CS}
\frac{d^2\sigma^{\nu(\bar{\nu})A}_{CC}}{dx_A\;dy_A}&=&\frac{G^2_FM_AE_\nu}{\pi}\left(\frac{M^2_W}{Q^2+M^2_W}\right)^2\Bigg[y^2_Ax_AF^{\nu(\bar{\nu})A}_{1}\nonumber\\
&&+\left(1-y_A-\frac{M_Ax_Ay_A}{2E_\nu}\right)F^{\nu(\bar{\nu})A}_{2} \nonumber \\
&&\pm~ x_Ay_A\left(1-\frac{y_A}{2}\right)F^{\nu(\bar{\nu})A}_{3}\Bigg] .\label{eq:CC_cross_section_xA_yA}
\end{eqnarray}
where $M_{A}$ is the mass of nucleus $A$. $F^{A}_{1}$, $F^{A}_{2}$ and $F^{A}_{3}$ are the structure functions for nucleus. The Bjorken variable in the nucleus is $x_{A}=x/A$, where $A$ is the mass number of the nucleus. The inelasticity in the nucleus is $y_{A}=y$~\cite{haider}.

  To correct for the nuclear effects in the structure functions $F^{A}_{2}$ and $F^{A}_{3}$,
we use EPPS16 package~\cite{Eskola:2016oht}. 
 EPPS16 is a package to obtain next-to-leading order (NLO) nuclear partonic distribution functions (nPDFs). The bound nucleon PDFs $f^{A}_{i} (x,Q^{2})$ for each parton flavor $i$ are given as:
\begin{equation}
f^{A}_{i} (x,Q^{2}) = R^{A}_{i} (x,Q^{2})~f^{CT10}_{i} (x,Q^{2}) .
\end{equation}
where $R^{A}_{i} (x,Q^{2})$ are the nuclear corrections to the free nucleon PDFs $f^{CT10}_{i} (x,Q^{2})$. EPPS16 provides parameterization only in the kinematical domain $1e-7 \leq x \leq 1$ and $1.3 \leq Q \leq 10000$ GeV.   

The neutrino structure function $F^{A}_{3}(x,Q^{2})$ on nucleus $A$ using EPPS16~\cite{Eskola:2016oht} is calculated  as:
\begin{eqnarray}
F^{A}_{3}(x,Q^{2})&=&u^{A}(x,Q) - \bar u^{A}(x,Q) + d^{A}(x,Q) \nonumber \\
&&-~\bar d^{A}(x,Q) + s^{A}(x,Q) -\bar s^{A}(x,Q)   \nonumber \\
&&+~c^{A}(x,Q)  - \bar c^{A}(x,Q) + b^{A}(x,Q) \nonumber \\
&&-~\bar b^{A}(x,Q) + t^{A}(x,Q)- \bar t^{A}(x,Q) . 
\end{eqnarray} 

The neutrino structure function $F^{A}_{2}(x,Q^{2})$ on nucleus $A$ using EPPS16~\cite{Eskola:2016oht} is calculated  as:
\begin{eqnarray}
F^{A}_{2}(x,Q^{2})&=&x~\Bigg(u^{A}(x,Q)+ \bar u^{A}(x,Q) + d^{A}(x,Q) \nonumber \\
&&+ ~\bar d^{A}(x,Q)+ s^{A}(x,Q) + \bar s^{A}(x,Q) \nonumber \\
&&+~c^{A}(x,Q) + \bar c^{A}(x,Q) + b^{A}(x,Q) \nonumber \\
&&+ ~\bar b^{A}(x,Q) + t^{A}(x,Q) + \bar t^{A}(x,Q)\Bigg) .
\end{eqnarray}

\subsection*{Target mass correction}
The target mass correction (TMC) can be taken into account when partonic distribution functions evaluate at Nachtmann variable $\xi$~\cite{nian} rather than Bjorken variable $x$ as:
\begin{equation}
\xi=\frac{2~x}{1 + \sqrt{1+\frac{4~M_{N}^{2}~x^{2}}{Q^{2}}}} .
\end{equation}
At high $Q^{2}(Q^{2} \gg M_{N}^{2})$, $\xi$ is equivalent to $x$.

\section{Results and discussions}

\begin{figure*}
\includegraphics[width=170mm]{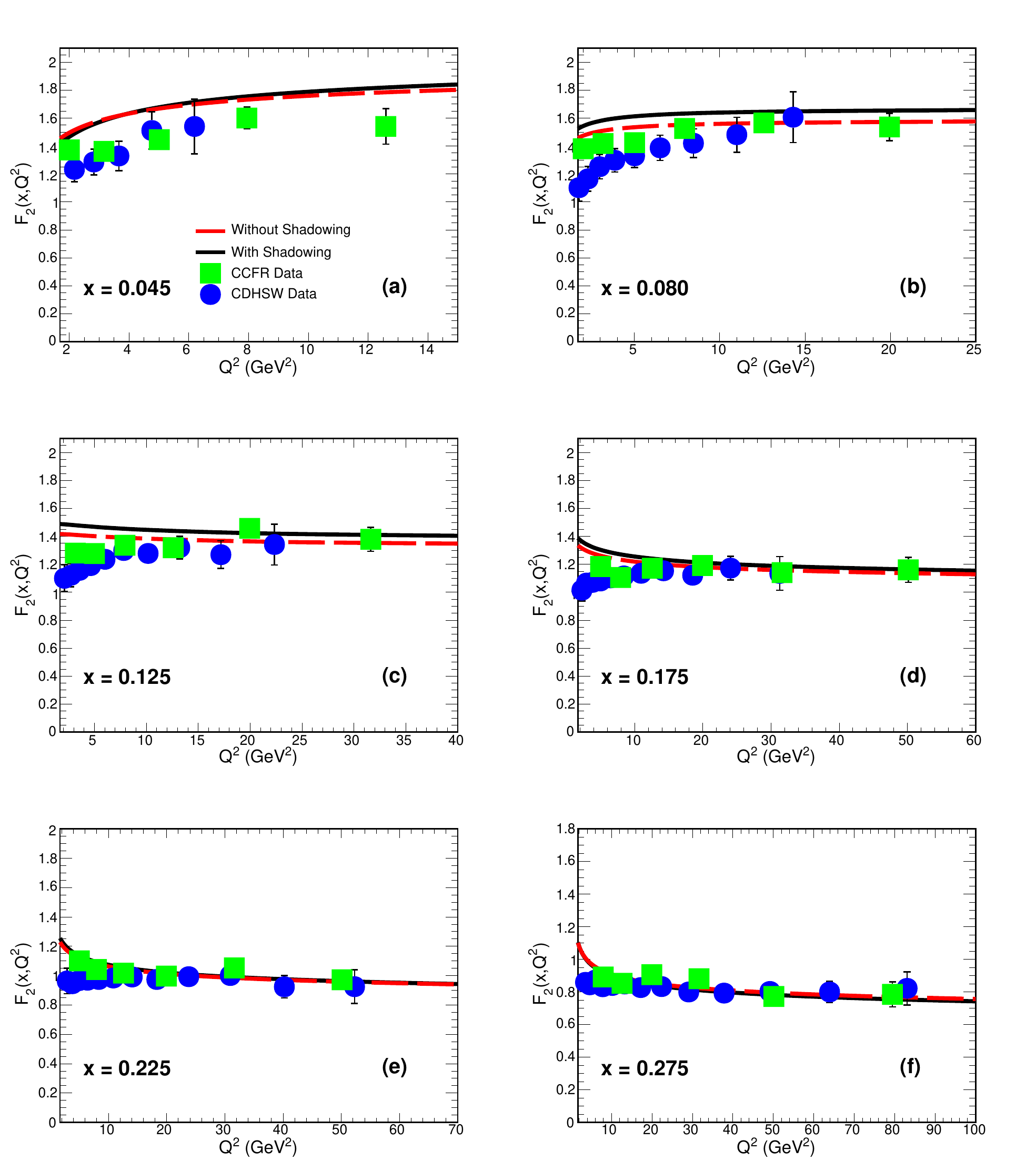}%
\caption{\label{F2_1} $F_{2}(x,Q^{2})$ with EPPS16 nuclear corrections~\cite{Eskola:2016oht} at NLO and LHAPDF (CT10) parton distribution functions~\cite{Lai:2010vv} for $^{56}$Fe as a function of $Q^{2}$ for different values of Bjorken variable $x$. The results are compared with measured data of CDHSW~\cite{Berge:1989hr} and CCFR~\cite{Oltman:1992pq} experiments.}
\end{figure*}

\begin{figure*}
\includegraphics[width=170mm]{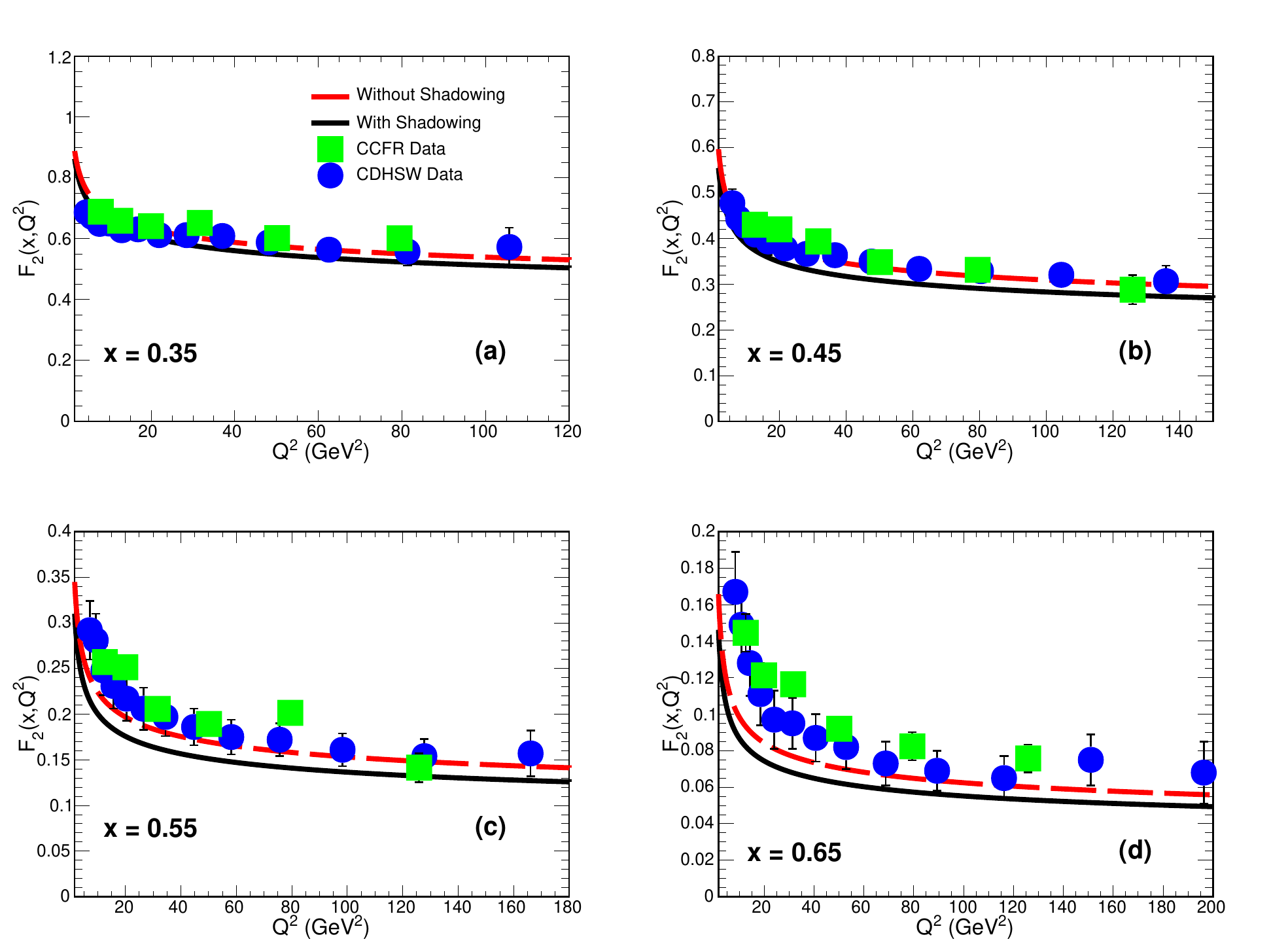}%
\caption{\label{F2_2} $F_{2}(x,Q^{2})$ with EPPS16 nuclear corrections~\cite{Eskola:2016oht} at NLO and LHAPDF (CT10) parton distribution functions~\cite{Lai:2010vv} for $^{56}$Fe as a function of $Q^{2}$ for different values of Bjorken variable $x$. The results are compared with measured data of CDHSW~\cite{Berge:1989hr} and CCFR~\cite{Oltman:1992pq} experiments.}
\end{figure*}

\begin{figure*}
\includegraphics[width=170mm]{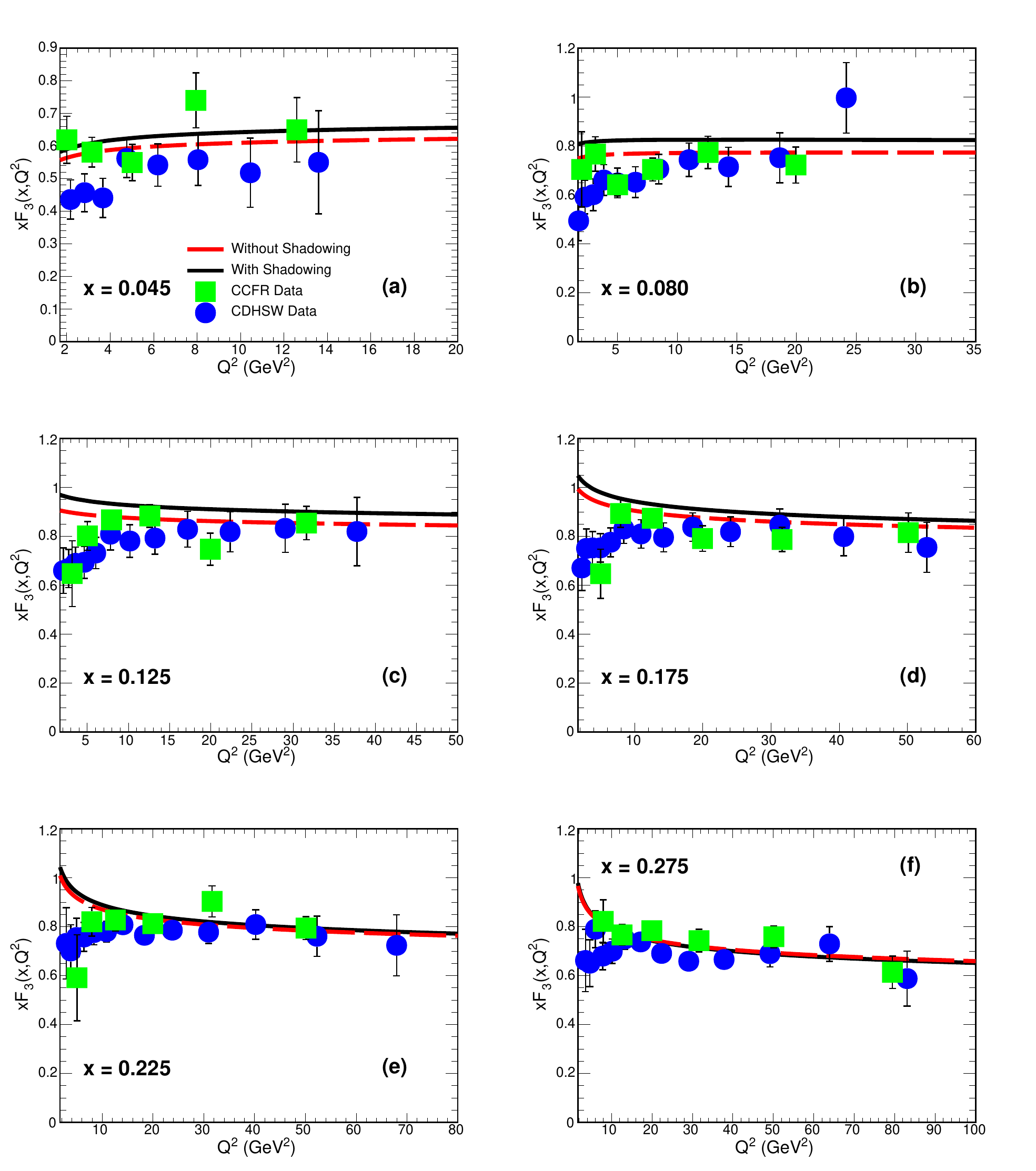}%
\caption{\label{xF3_1} $xF_{3}(x,Q^{2})$ with EPPS16 nuclear corrections~\cite{Eskola:2016oht} at NLO and LHAPDF (CT10) parton distribution functions~\cite{Lai:2010vv} for $^{56}$Fe as a function of $Q^{2}$ for different values of Bjorken variable $x$. The results are compared with measured data of CDHSW~\cite{Berge:1989hr} and CCFR~\cite{Oltman:1992pq} experiments.}
\end{figure*}

\begin{figure*}
\includegraphics[width=170mm]{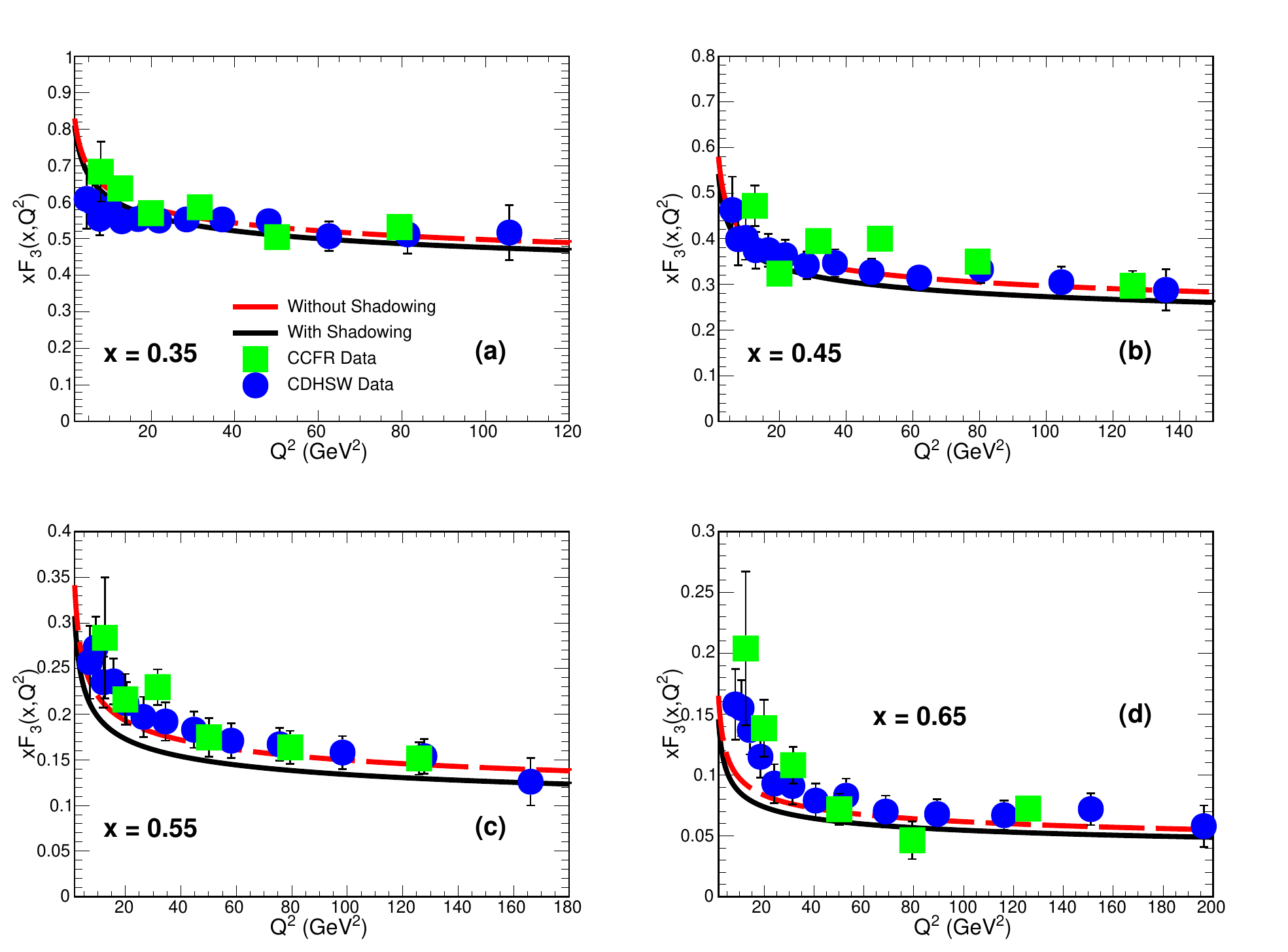}%
\caption{\label{xF3_2} $xF_{3}(x,Q^{2})$ with EPPS16 nuclear corrections~\cite{Eskola:2016oht} at NLO and LHAPDF (CT10) parton distribution functions~\cite{Lai:2010vv} for $^{56}$Fe as a function of $Q^{2}$ for different values of Bjorken variable $x$. The results are compared with measured data of CDHSW~\cite{Berge:1989hr} and CCFR~\cite{Oltman:1992pq} experiments.}
\end{figure*}

 We calculated the structure functions $F_{2}(x,Q^{2})$ and $xF_{3}(x,Q^{2})$  for $^{56}$Fe with EPPS16 nuclear corrections~\cite{Eskola:2016oht} at next-to-leading order (NLO) and LHAPDF (CT10) parton distribution functions~\cite{Lai:2010vv}. Figures \ref{F2_1} and \ref{F2_2} show present calculations of $F_{2}(x,Q^{2})$ as a function of the square of momentum transfer $Q^{2}$, for different values of Bjorken variable $x$ (0.045, 0.080, 0.125, 0.175, 0.225, 0.275, 0.35, 0.45, 0.55, 0.65). Figures \ref{xF3_1} and \ref{xF3_2} show present calculations of the structure function $xF_{3}(x,Q^{2})$ as a function of the square of momentum transfer $Q^{2}$, for different values of Bjorken variable $x$. In figures \ref{F2_1} and \ref{xF3_1}, panel (a) is for $x = 0.045$, panel (b) is for $x = 0.080$, panel (c) is for $x = 0.125$, panel (d) is for $x = 0.175$, panel (e) is for $x = 0.225$ and panel (f) is for $x = 0.275$. In figures \ref{F2_2} and \ref{xF3_2}, panel (a) is for $x = 0.35$, panel (b) is for $x = 0.45$, panel (c) is for $x = 0.55$ and panel (d) is for $x = 0.65$. The black solid lines show the calculations with the inclusion of shadowing effect where as the red dashed lines show the calculations without the inclusion of shadowing effect. The obtained results are compared with measured data of CDHSW~\cite{Berge:1989hr} and CCFR~\cite{Oltman:1992pq} experiments. We can see that for higher values of Bjorken variable $x$, the present theoretical approach gives a good description of data. Figure \ref{F2_2}(a) and figure \ref{xF3_2}(a) show an excellent agreement between theoretical calculations and data for $x = 0.35$.

\begin{figure*}
\begin{tabular}{cc}
\includegraphics[width=85mm]{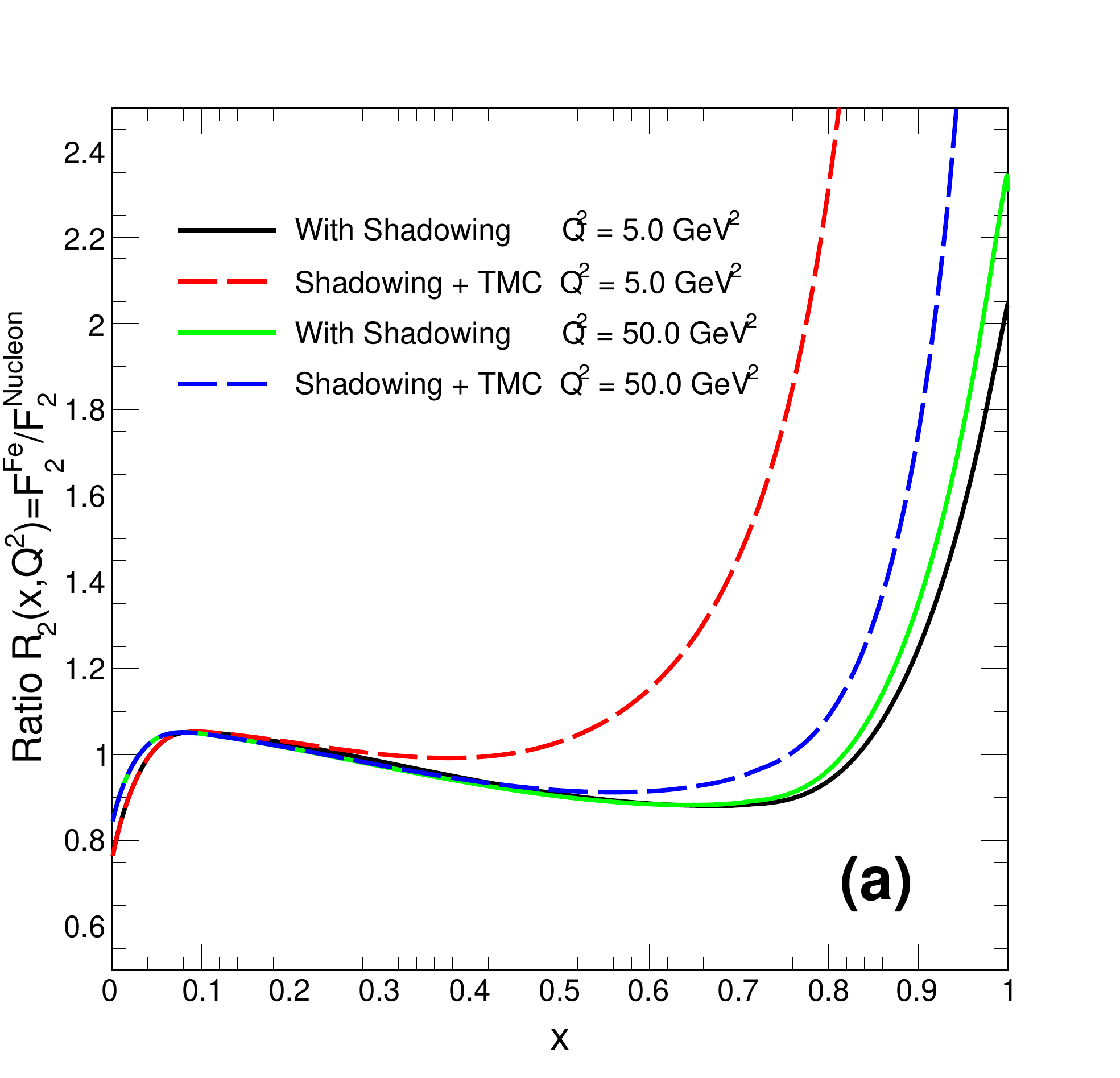}&
\includegraphics[width=85mm]{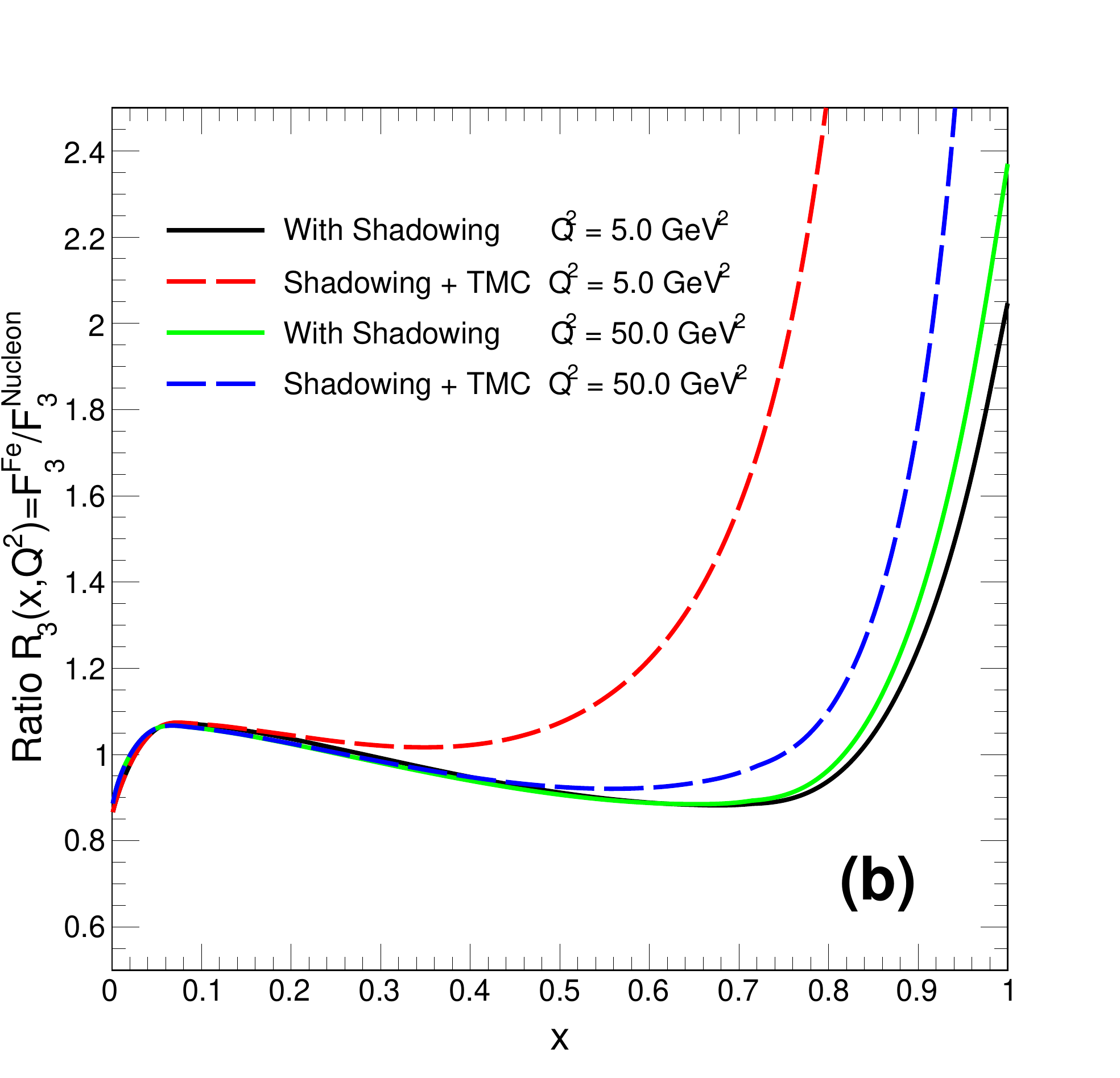}\\
\end{tabular}
\caption{\label{R2_R3} Ratios (a) $R_{2}(x,Q^{2}) = \frac{F^{^{56}Fe}_{2}}{F^{Nucleon}_{2}}$ and (b) $R_{3}(x,Q^{2}) = \frac{F^{^{56}Fe}_{3}}{F^{Nucleon}_{3}}$ as a function of Bjorken variable $x$ with EPPS16 nuclear corrections~\cite{Eskola:2016oht} at NLO and LHAPDF (CT10) parton distribution functions~\cite{Lai:2010vv} for $Q^{2} = 5.0$ GeV$^{2}$ and $Q^{2} = 50.0$ GeV$^{2}$.}
\end{figure*}

Figure \ref{R2_R3}(a) shows present calculations of ratio $R_{2}(x,Q^{2}) = \frac{F^{^{56}Fe}_{2}}{F^{Nucleon}_{2}}$ as a function of Bjorken variable $x$ for $Q^{2} = 5.0$ GeV$^{2}$ and $Q^{2} = 50.0$ GeV$^{2}$. Figure \ref{R2_R3}(b) shows present calculations of ratio $R_{3}(x,Q^{2}) = \frac{F^{^{56}Fe}_{3}}{F^{Nucleon}_{3}}$ as a function of Bjorken variable $x$ for $Q^{2} = 5.0$ GeV$^{2}$ and $Q^{2} = 50.0$ GeV$^{2}$. The broken lines show the effect of target mass correction~\cite{nian}. We can clearly see the effects discussed in section \ref{section_nuc_modifications}.

\begin{figure*}
\includegraphics[width=170mm]{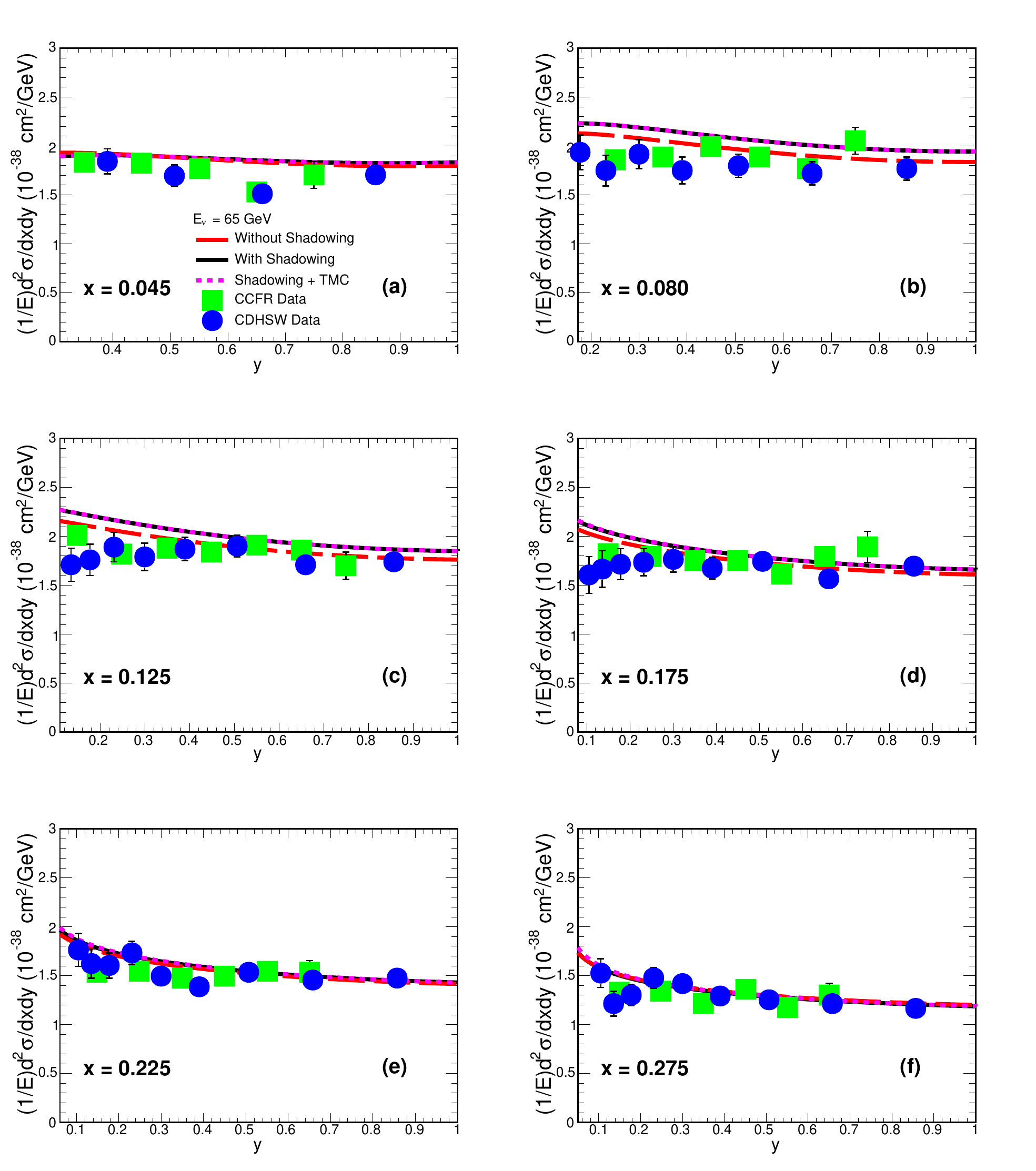}%
\caption{\label{dsigma_65GeV_1} $\frac{1}{E}\frac{d^{2}\sigma}{dxdy}$ for $^{56}$Fe as a function of inelasticity $y$, for different values of Bjorken variable $x$ and $E_{\nu_{\mu}} = 65$ GeV using EPPS16 nuclear corrections~\cite{Eskola:2016oht} at NLO and LHAPDF (CT10) parton distribution functions~\cite{Lai:2010vv}. Results are compared with measured data of CCFR~\cite{Yang:2001rm} and CDHSW~\cite{Yang:2001rm} experiments.}
\end{figure*}

\begin{figure*}
\includegraphics[width=170mm]{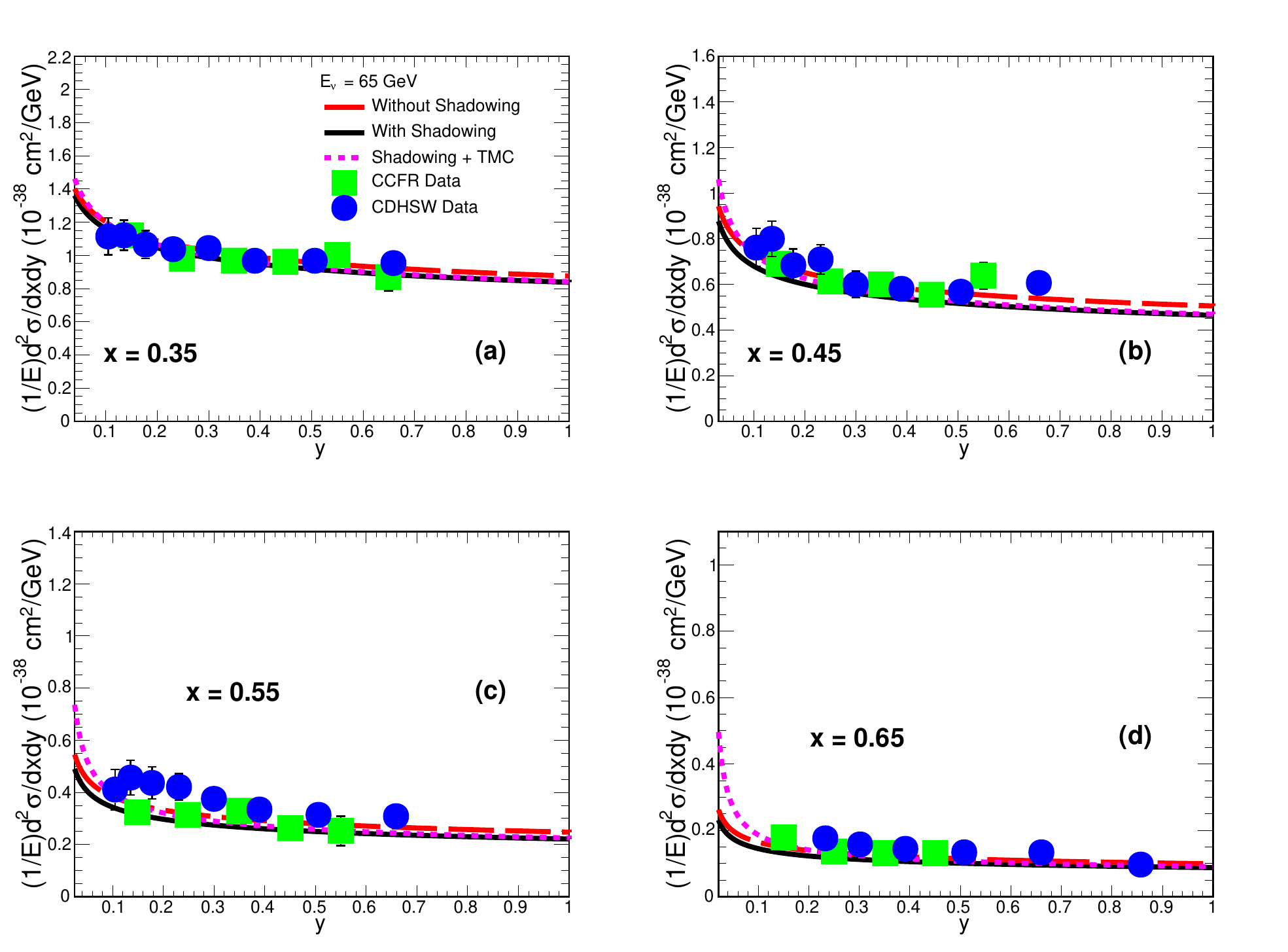}%
\caption{\label{dsigma_65GeV_2} $\frac{1}{E}\frac{d^{2}\sigma}{dxdy}$ for $^{56}$Fe as a function of inelasticity $y$, for different values of Bjorken variable $x$ and $E_{\nu_{\mu}} = 65$ GeV using EPPS16 nuclear corrections~\cite{Eskola:2016oht} at NLO and LHAPDF (CT10) parton distribution functions~\cite{Lai:2010vv}. Results are compared with measured data of CCFR~\cite{Yang:2001rm} and CDHSW~\cite{Yang:2001rm} experiments.}
\end{figure*}

\begin{figure*}
\includegraphics[width=170mm]{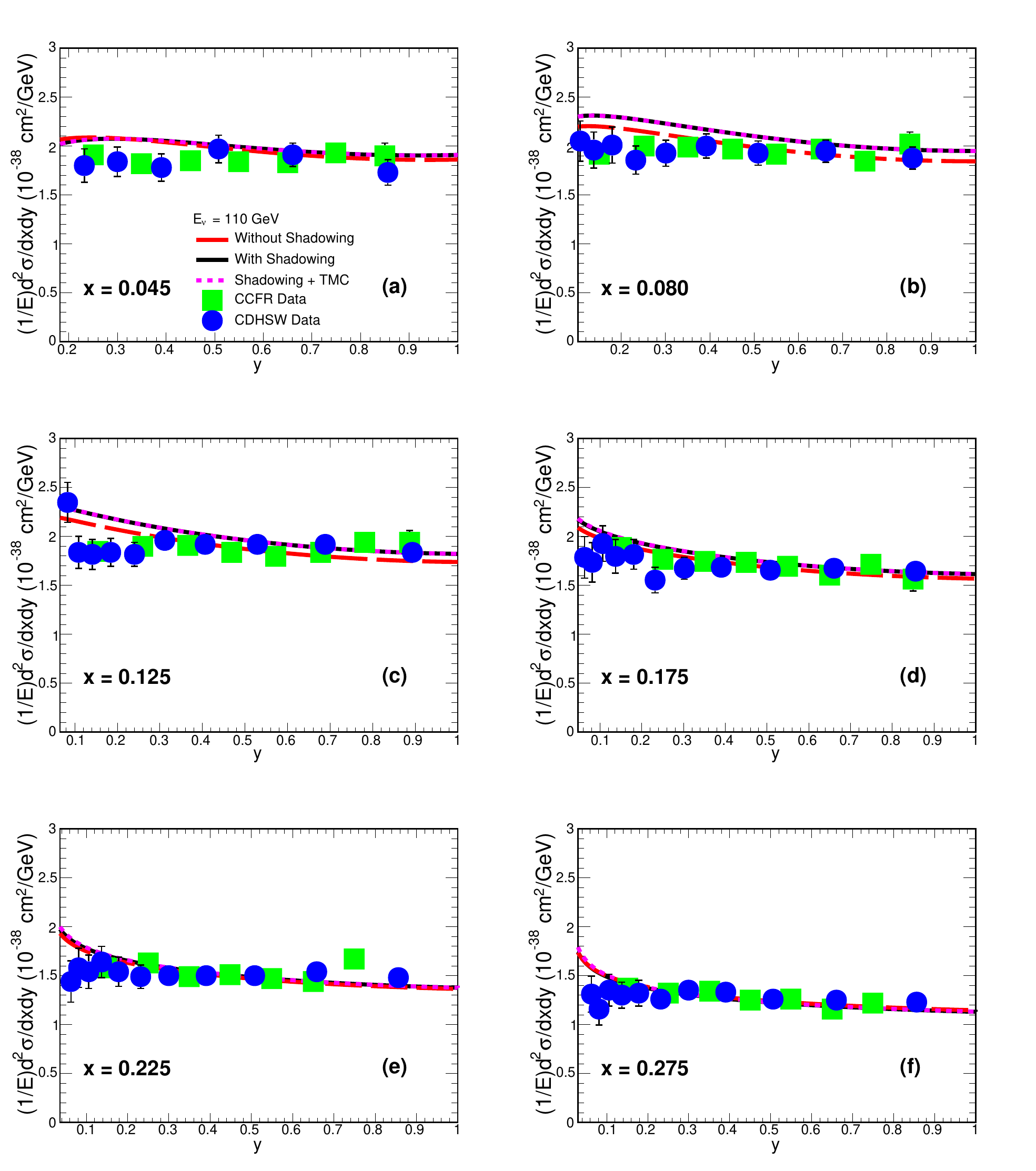}%
\caption{\label{dsigma_110GeV_1} $\frac{1}{E}\frac{d^{2}\sigma}{dxdy}$ for $^{56}$Fe as a function of inelasticity $y$, for different values of Bjorken variable $x$ and $E_{\nu_{\mu}} = 110$ GeV using EPPS16 nuclear corrections~\cite{Eskola:2016oht} at NLO and LHAPDF (CT10) parton distribution functions~\cite{Lai:2010vv}. Results are compared with measured data of CCFR~\cite{Yang:2001rm} and CDHSW~\cite{Yang:2001rm} experiments.}
\end{figure*}

\begin{figure*}
\includegraphics[width=170mm]{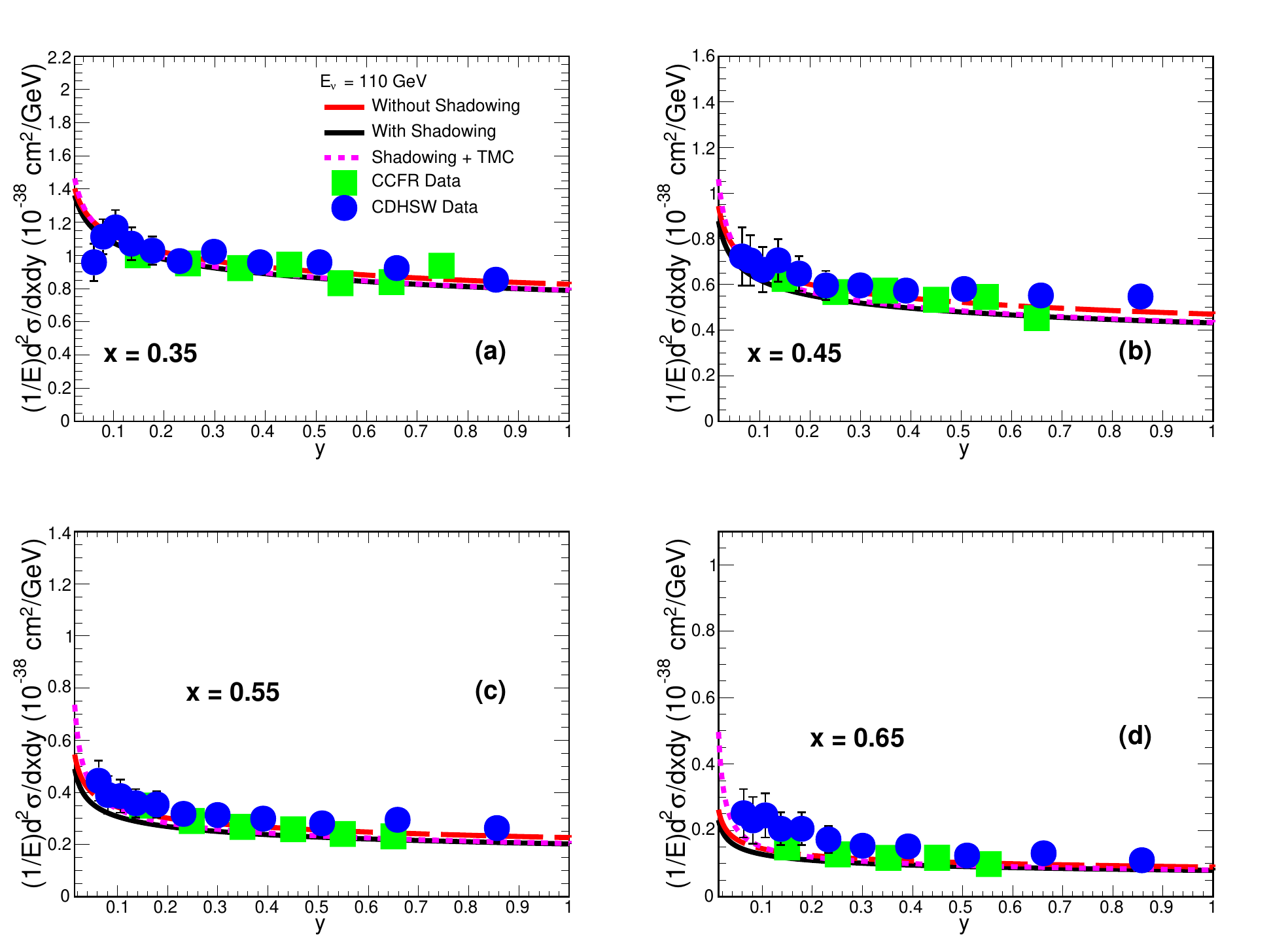}%
\caption{\label{dsigma_110GeV_2} $\frac{1}{E}\frac{d^{2}\sigma}{dxdy}$ for $^{56}$Fe as a function of inelasticity $y$, for different values of Bjorken variable $x$ and $E_{\nu_{\mu}} = 110$ GeV using EPPS16 nuclear corrections~\cite{Eskola:2016oht} at NLO and LHAPDF (CT10) parton distribution functions~\cite{Lai:2010vv}. Results are compared with measured data of CCFR~\cite{Yang:2001rm} and CDHSW~\cite{Yang:2001rm} experiments.}
\end{figure*}

\begin{figure*}
\includegraphics[width=170mm]{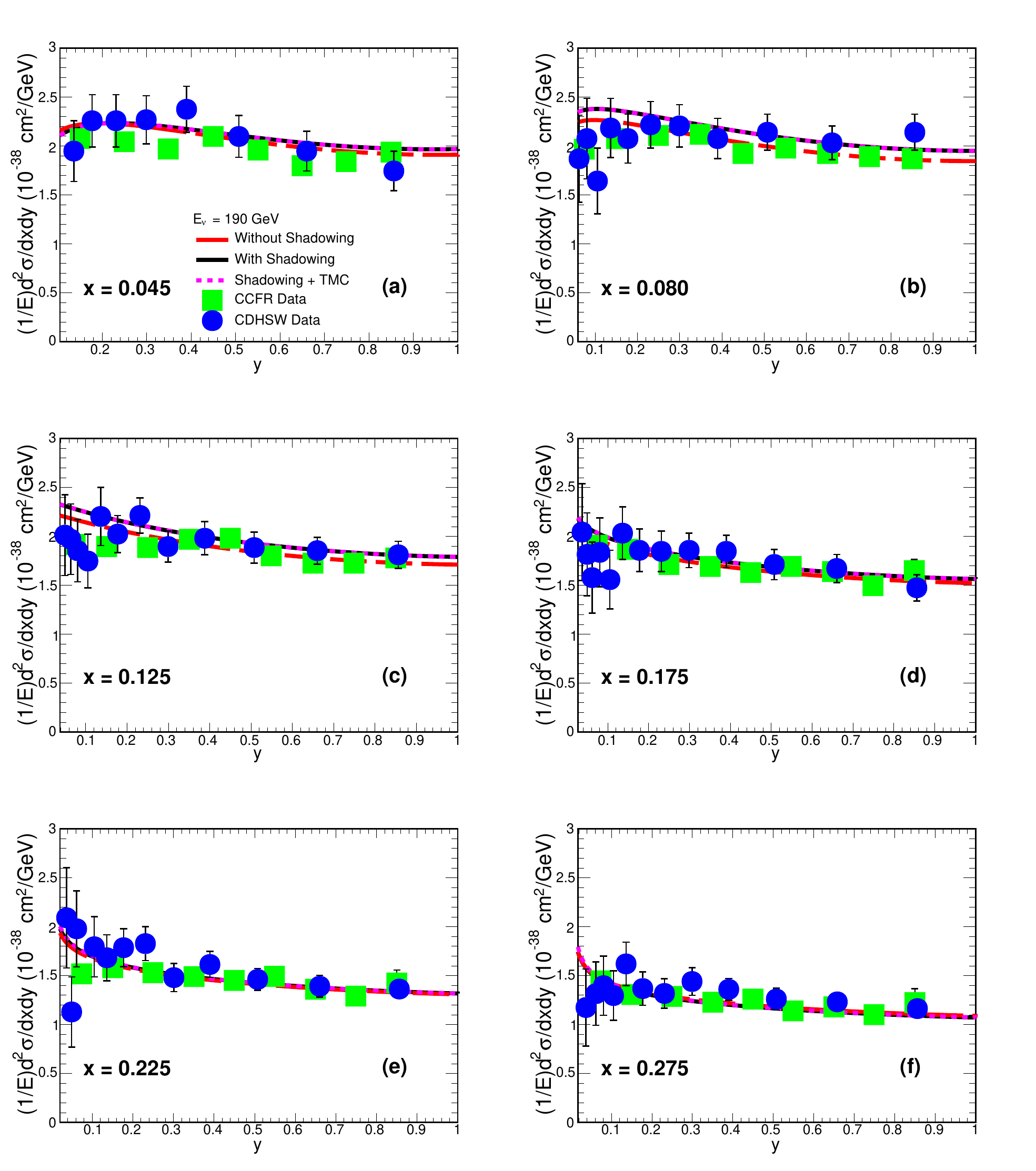}%
\caption{\label{dsigma_190GeV_1} $\frac{1}{E}\frac{d^{2}\sigma}{dxdy}$ for $^{56}$Fe as a function of inelasticity $y$, for different values of Bjorken variable $x$ and $E_{\nu_{\mu}} = 190$ GeV using EPPS16 nuclear corrections~\cite{Eskola:2016oht} at NLO and LHAPDF (CT10) parton distribution functions~\cite{Lai:2010vv}. Results are compared with measured data of CCFR~\cite{Yang:2001rm} and CDHSW~\cite{Yang:2001rm} experiments.}
\end{figure*}

\begin{figure*}
\includegraphics[width=170mm]{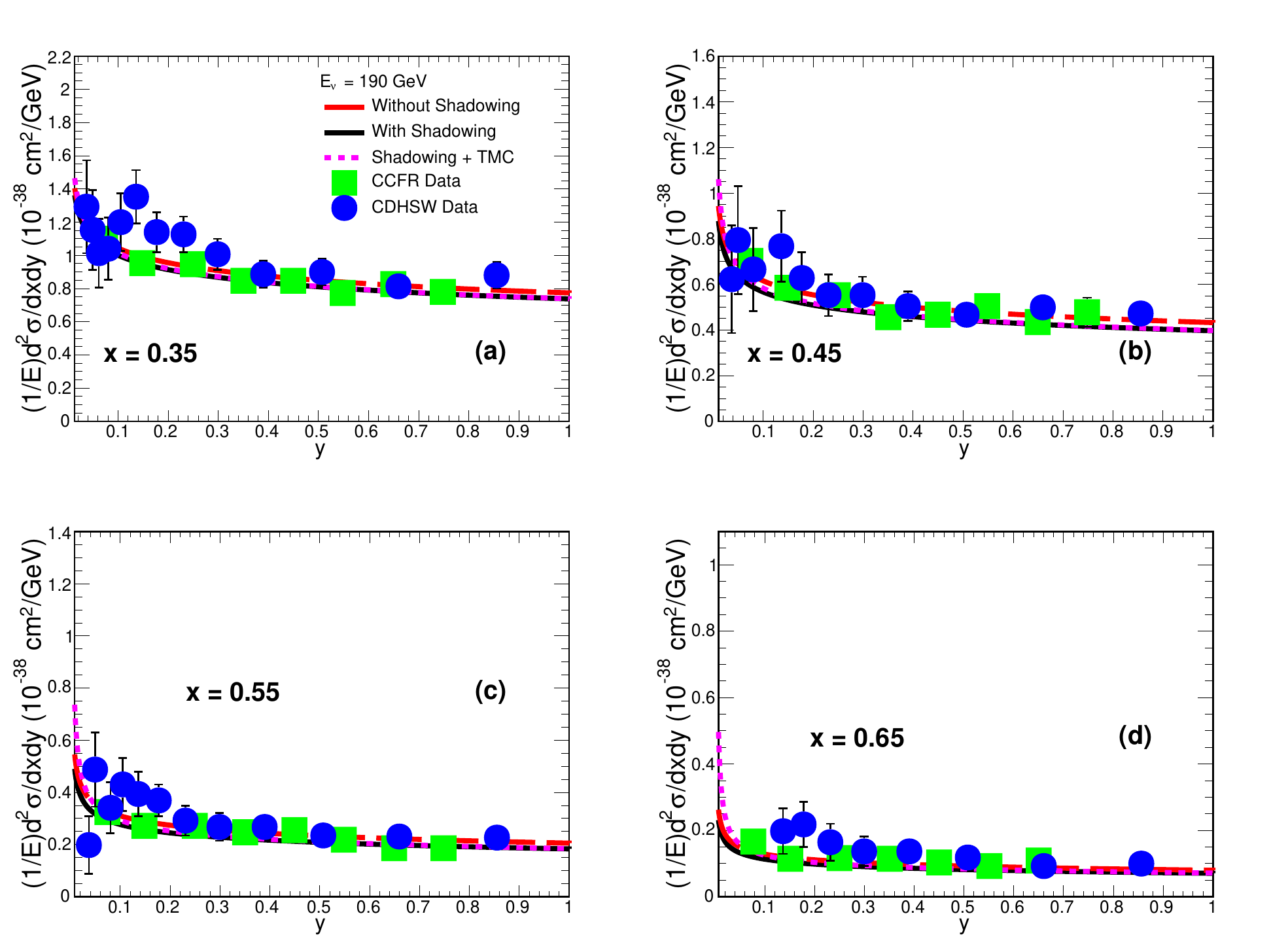}%
\caption{\label{dsigma_190GeV_2} $\frac{1}{E}\frac{d^{2}\sigma}{dxdy}$ for $^{56}$Fe as a function of inelasticity $y$, for different values of Bjorken variable $x$ and $E_{\nu_{\mu}} = 190$ GeV using EPPS16 nuclear corrections~\cite{Eskola:2016oht} at NLO and LHAPDF (CT10) parton distribution functions~\cite{Lai:2010vv}. Results are compared with measured data of CCFR~\cite{Yang:2001rm} and CDHSW~\cite{Yang:2001rm} experiments.}
\end{figure*}

We calculated $\frac{1}{E}\frac{d^{2}\sigma}{dxdy}$ for $^{56}$Fe as a function of inelasticity $y$, for different values of Bjorken variable $x$ and for different neutrino energies. Figures \ref{dsigma_65GeV_1} and \ref{dsigma_65GeV_2} show present calculations of $\frac{1}{E}\frac{d^{2}\sigma}{dxdy}$ for $E_{\nu_{\mu}} = 65$ GeV. Figures \ref{dsigma_110GeV_1} and \ref{dsigma_110GeV_2} show the calculations of $\frac{1}{E}\frac{d^{2}\sigma}{dxdy}$ for $E_{\nu_{\mu}} = 110$ GeV. Figures \ref{dsigma_190GeV_1} and \ref{dsigma_190GeV_2} show the similar calculations of $\frac{1}{E}\frac{d^{2}\sigma}{dxdy}$ for $E_{\nu_{\mu}} = 190$ GeV. The results obtained are compared with measured data of CCFR~\cite{Yang:2001rm} and CDHSW~\cite{Yang:2001rm} experiments. In figures \ref{dsigma_65GeV_1}, \ref{dsigma_110GeV_1} and \ref{dsigma_190GeV_1}, panel (a) is for $x = 0.045$, panel (b) is for $x = 0.080$, panel (c) is for $x = 0.125$, panel (d) is for $x = 0.175$, panel (e) is for $x = 0.225$ and panel (f) is for $x = 0.275$. In figures \ref{dsigma_65GeV_2}, \ref{dsigma_110GeV_2} and \ref{dsigma_190GeV_2}, panel (a) is for $x = 0.35$, panel (b) is for $x = 0.45$, panel (c) is for $x = 0.55$ and panel (d) is for $x = 0.65$. The black solid lines show the calculations with the inclusion of shadowing effect, the red dashed lines show the calculations without the inclusion of shadowing effect and the pink dotted lines show the calculations with the inclusion of shadowing effect and target mass correction.

\clearpage

\noindent
We can see that the target mass correction has negligible effect at low values of Bjorken variable $x$, but the effect increases with the increasing values of $x$. Also TMC effects are seen for low values of inelasticity $y$. Here also we can see that with the increasing values of $x$, theoretical calculations show a better agreement with data.

\section{Conclusion}

 We presented the structure functions ($F_{2}(x,Q^{2})$ and $xF_{3}(x,Q^{2})$), the ratios ($R_{2}(x,Q^{2}) = \frac{F^{^{56}Fe}_{2}}{F^{Nucleon}_{2}}$ and $R_{3}(x,Q^{2}) = \frac{F^{^{56}Fe}_{3}}{F^{Nucleon}_{3}}$) and the differential cross sections for charged current $\nu_{\mu}$ - nucleon and $\nu_{\mu}$ - $^{56}$Fe deep inelastic scattering using a formalism based on Hirai, Kumao and Saito model. We used LHAPDF (CT10) parton distribution functions. Nuclear corrections inside the nuclei are applied to the PDFs at next-to-leading order using EPPS16 parameterization. We also incorporated target mass correction to our calculations. We studied the behavior of the structure functions $F_{2}(x,Q^{2})$ and $xF_{3}(x,Q^{2})$ as a function of the square of momentum transfer $Q^{2}$ for different values of Bjorken variable $x$. Differential cross sections are analyzed as a function of inelasticity $y$ for different values of Bjorken variable $x$ and for different neutrino energies. The results obtained have been compared with measured experimental data. The present theoretical approach gives a good description of data. The agreement between theoretical calculations and data is even better for higher values of Bjorken variable $x$.

\begin{acknowledgments}
The authors are thankful to Department of Science and Technology (DST), New Delhi, Government of India for financial support needed to pursue this work.
\end{acknowledgments}



\noindent
\begin{thebibliography}{100}
\medskip
\bibitem{Ahn:2002up} 
  M.~H.~Ahn {\it et al.} [K2K Collaboration],
  Phys.\ Rev.\ Lett.\  {\bf 90}, 041801 (2003).
\bibitem{Aliu:2004sq} 
  E.~Aliu {\it et al.} [K2K Collaboration],
  Phys.\ Rev.\ Lett.\  {\bf 94}, 081802 (2005).
\bibitem{Ahn:2006zza} 
  M.~H.~Ahn {\it et al.} [K2K Collaboration],
  Phys.\ Rev.\ D {\bf 74}, 072003 (2006).
\bibitem{Ashie:2005ik} 
  Y.~Ashie {\it et al.} [Super-Kamiokande Collaboration],
  Phys.\ Rev.\ D {\bf 71}, 112005 (2005).
\bibitem{Takeuchi:2011aa} 
  Y.~Takeuchi [Super-Kamiokande Collaboration],
  Nucl.\ Phys.\ Proc.\ Supl.\  {\bf 229-232}, 79 (2012).
\bibitem{Adamson:2016tbq} 
  P.~Adamson {\it et al.} [NOvA Collaboration],
  Phys.\ Rev.\ Lett.\  {\bf 116}, no. 15, 151806 (2016).
\bibitem{Adamson:2016xxw} 
  P.~Adamson {\it et al.} [NOvA Collaboration],
  Phys.\ Rev.\ D {\bf 93}, no. 5, 051104 (2016).
\bibitem{Adamson:2017qqn} 
  P.~Adamson {\it et al.} [NOvA Collaboration],
  Phys.\ Rev.\ Lett.\  {\bf 118}, no. 15, 151802 (2017).
\bibitem{Adamson:2017gxd} 
  P.~Adamson {\it et al.} [NOvA Collaboration],
  Phys.\ Rev.\ Lett.\  {\bf 118}, no. 23, 231801 (2017).
\bibitem{Kumar:2017sdq} 
  S.~Ahmed {\it et al.} [ICAL Collaboration],
  Pramana {\bf 88}, no. 5, 79 (2017).
\bibitem{Grover:2018out} 
  D.~Grover, K.~Saraswat, P.~Shukla and V.~Singh,
  arXiv:1807.08911 [hep-ph].
\bibitem{Saraswat:2016kln} 
  K.~Saraswat, P.~Shukla, V.~Kumar and V.~Singh,
  Phys.\ Rev.\ C {\bf 93}, 035504 (2016).
\bibitem{Formaggio:2013kya} 
  J.~A.~Formaggio and G.~P.~Zeller,
  Rev.\ Mod.\ Phys.\  {\bf 84}, 1307 (2012).
\bibitem{Lipari:2002at} 
  P.~Lipari,
  Nucl.\ Phys.\ Proc.\ Suppl.\  {\bf 112}, 274 (2002).
\bibitem{Tzanov:2005kr} 
  M.~Tzanov {\it et al.} [NuTeV Collaboration],
  Phys.\ Rev.\ D {\bf 74}, 012008 (2006).
\bibitem{Onengut:2005kv} 
  G.~Onengut {\it et al.} [CHORUS Collaboration],
  Phys.\ Lett.\ B {\bf 632}, 65 (2006).
\bibitem{Wu:2007ab} 
  Q.~Wu {\it et al.} [NOMAD Collaboration],
  Phys.\ Lett.\ B {\bf 660}, 19 (2008).
\bibitem{Bhattacharya} 
  D.~Bhattacharya, Ph.D. thesis, University of Pittsburgh (2009).
\bibitem{Mousseau:2016snl} 
  J.~Mousseau {\it et al.} [MINERvA Collaboration],
  Phys.\ Rev.\ D {\bf 93}, no. 7, 071101 (2016).
\bibitem{Tzanov:2009zz} 
  M.~Tzanov,
  AIP Conf.\ Proc.\  {\bf 1222}, 243 (2010).
\bibitem{Martin:2009iq} 
  A.~D.~Martin, W.~J.~Stirling, R.~S.~Thorne and G.~Watt,
  Eur.\ Phys.\ J.\ C {\bf 63}, 189 (2009).
\bibitem{Martin:2009bu} 
  A.~D.~Martin, W.~J.~Stirling, R.~S.~Thorne and G.~Watt,
  Eur.\ Phys.\ J.\ C {\bf 64}, 653 (2009).
\bibitem{Martin:2010db} 
  A.~D.~Martin, W.~J.~Stirling, R.~S.~Thorne and G.~Watt,
  Eur.\ Phys.\ J.\ C {\bf 70}, 51 (2010).
\bibitem{Nadolsky:2008zw} 
  P.~M.~Nadolsky, H.~L.~Lai, Q.~H.~Cao, J.~Huston, J.~Pumplin, D.~Stump, W.~K.~Tung and C.-P.~Yuan,
  Phys.\ Rev.\ D {\bf 78}, 013004 (2008).
\bibitem{Alekhin:2013nda} 
  S.~Alekhin, J.~Blumlein and S.~Moch,
  Phys.\ Rev.\ D {\bf 89}, no. 5, 054028 (2014).
\bibitem{Alekhin:2015cza} 
  S.~Alekhin, J.~Blümlein, S.~Moch and R.~Placakyte,
  Phys.\ Rev.\ D {\bf 94}, no. 11, 114038 (2016).
\bibitem{Chekanov:2002pv} 
  S.~Chekanov {\it et al.} [ZEUS Collaboration],
  Phys.\ Rev.\ D {\bf 67}, 012007 (2003).
\bibitem{whalley} 
  M.~R.~Whalley, D.~Bourilkov and R.~C.~Group,
  hep-ph/0508110.
\bibitem{Lai:2010vv} 
  H.~L.~Lai, M.~Guzzi, J.~Huston, Z.~Li, P.~M.~Nadolsky, J.~Pumplin and C.-P.~Yuan,
  Phys.\ Rev.\ D {\bf 82}, 074024 (2010).
\bibitem{hirai} 
  M.~Hirai, S.~Kumano and K.~Saito,
  AIP Conf.\ Proc.\  {\bf 1189}, 269 (2009).
\bibitem{callan} 
  C.~G.~Callan, Jr. and D.~J.~Gross,
  Phys.\ Rev.\ Lett.\  {\bf 22}, 156 (1969).
\bibitem{Armesto:2006ph} 
  N.~Armesto,
  J.\ Phys.\ G {\bf 32}, R367 (2006).
\bibitem{Aubert:1983xm} 
  J.~J.~Aubert {\it et al.} [European Muon Collaboration],
  Phys.\ Lett.\  {\bf 123B}, 275 (1983).
\bibitem{Arneodo:1992wf} 
  M.~Arneodo,
  Phys.\ Rept.\  {\bf 240}, 301 (1994).
\bibitem{haider} 
  H.~Haider, I.~Ruiz Simo and M.~Sajjad Athar,
  Phys.\ Rev.\ C {\bf 85}, 055201 (2012).
\bibitem{Eskola:2016oht} 
  K.~J.~Eskola, P.~Paakkinen, H.~Paukkunen and C.~A.~Salgado,
  Eur.\ Phys.\ J.\ C {\bf 77}, no. 3, 163 (2017).
\bibitem{nian} 
  D.~ChunGui, L.~GuangLie and S.~PengNian,
  Eur.\ Phys.\ J.\ C {\bf 48}, 125 (2006).
\bibitem{Berge:1989hr} 
  J.~P.~Berge {\it et al.},
  Z.\ Phys.\ C {\bf 49}, 187 (1991).
\bibitem{Oltman:1992pq} 
  E.~Oltman {\it et al.},
  Z.\ Phys.\ C {\bf 53}, 51 (1992).
\bibitem{Yang:2001rm} 
  U.~K.~Yang,
  FERMILAB-THESIS-2001-09, UMI-99-98273.
\end{thebibliography}
\end{document}